\documentclass[11pt]{article}
\usepackage{amssymb}
\usepackage{amsthm}
\usepackage{amsmath,amsfonts,amssymb}
\usepackage{setspace}

\usepackage[mathscr]{eucal}
\usepackage{exscale}
\usepackage{natbib}
\usepackage{bm}
\usepackage{eqlist}
\usepackage[dvipsnames]{color}
\usepackage[Lenny]{fncychap}
\usepackage{multirow}
\usepackage{graphicx}
\usepackage{algpseudocode}
\usepackage{algorithm}
\usepackage{caption}
\usepackage{hyperref}
\usepackage{framed}
\usepackage{color}
\usepackage{booktabs}
\usepackage{makecell}
\usepackage{longtable}

\usepackage{framed}
\usepackage{algpseudocode}

\usepackage{tabu}
\usepackage{tikz}

\usepackage[utf8]{inputenc} 
\usepackage[english]{babel} 
\usepackage[T1]{fontenc}    
\usepackage{amssymb,amsmath,amsfonts} 
\usepackage{braket} 
\usepackage{caption}
\usepackage{algorithm}

\hypersetup{
    colorlinks=true,
    linkcolor=blue,
    citecolor=blue,
    citebordercolor={1 1 0},
}

\newtheorem{theorem}{Theorem}[section]

\newtheorem{remark}[theorem]{Remark}

\font\bigbf=cmbx10 scaled \magstep3

\begin{document}

\title{\bigbf  A survey of urban drive-by sensing: An optimization perspective}

\author{Wen Ji
\quad 
Ke Han$\thanks{Corresponding author, e-mail: kehan@swjtu.edu.cn;}$
\quad 
Tao Liu
\\\\
\textit{\small Institute of System Science and Engineering, School of Transportation and Logistics,}\\
\textit{\small Southwest Jiaotong University, Chengdu, China, 611756}
}

\maketitle

\begin{abstract}
Pervasive and mobile sensing is an integral part of smart transport and smart city applications. Vehicle-based mobile sensing, or {\it drive-by sensing} (DS), is gaining popularity in both academic research and field practice. The DS paradigm has an inherent transport component, as the spatial-temporal distribution of the sensors are closely related to the mobility patterns of their hosts, which may include third-party (e.g. taxis, buses) or for-hire (e.g. unmanned aerial vehicles and dedicated vehicles) vehicles. It is therefore essential to understand, assess and optimize the sensing power of vehicle fleets under a wide range of urban sensing scenarios. To this end, this paper offers an optimization-oriented summary of recent literature by presenting a four-step discussion, namely (1) quantifying the sensing quality (objective); (2) assessing the sensing power of various fleets (strategic); (3) sensor deployment (strategic/tactical); and (4) vehicle maneuvers (tactical/operational). By compiling research findings and practical insights in this way, this review article not only highlights the optimization aspect of drive-by sensing, but also serves as a practical guide for configuring and deploying vehicle-based urban sensing systems.
\end{abstract}


\section{Introduction}\label{secIntro}

Central to many smart city applications is the capability to sense and monitor a variety of entities ranging from natural and built environment (e.g. through ubiquitous and pervasive sensing) to domestic and industrial processes (e.g. through IoT technologies) \citep{Chan2021,DSXVF2019,ZBCVZ2014}. Both cases benefit from recent development of wireless technology and miniaturized sensing apparatus \citep{PZCG2014,LSZ2016}. When it comes to the sensing of outdoor urban environment, such as air quality and traffic states, the uprise of mobile sensors has enabled extensive spatial-temporal coverage by leveraging human mobility in a cost-effective way, giving rise to a rapidly expanding literature on mobile crowdsensing \citep{GYL2011,LBDNT2005,MZY2014,ZWXG2014}. For a general review of the architecture, technology, and applications of mobile crowdsensing, the reader is referred to \cite{YLSLML2016, CFKFKB2019} and \cite{LSBC2020}.

According to the type of sensor hosts, mobile crowdsensing can be categorized as community sensing (CS) and drive-by sensing (DS), where the former utilizes sensors embedded in smartphones or wearable devices \citep{JSLTPYMLDH2011,DKSHHW2014,YXFT2012,SM2019}, and the latter relies on road or airborne vehicles such as taxis, buses, trams and unmanned aerial vehicles (UAVs) \citep{LG2010,Sanchez2014}. In general, DS outperforms CS in terms of geographic extent, temporal duration, reliability, cost-effectiveness and data quality \citep{YLMLLM2015}, and has been widely seen in various smart city applications such as air quality sensing \citep{KMMBNDBNB2015, HSWHFABT2015, SHS2021, Mahajan2021,Marjovi2015}, traffic conditions estimation \citep{GQDRJ2022,LSLHLW2009,YNL2007}, urban heat island phenomenon \citep{FBRDRAP2021}, noise monitoring \citep{Alsina-Pages2017}, and infrastructure health inspection \citep{WBS2014, Eriksson2008, MCG2022}.

In drive-by sensing, the sensors may be carried by taxis \citep{Honicky2008,Mathur2010}, transit vehicles such as buses or trams \citep{KPWLZW2016,Saukh2012,GCGGD2016}, service vehicles such as trash trucks \citep{ADRMdR2018, dADKKR2020}, dedicated vehicles \citep{Messier2018,VonFischer2017}, and UAVs \citep{HBYZBS2019,LLHWXWLZD2022,Hemamalini2022}. Clearly, the spatial-temporal distribution of sensors, and subsequently their sensing performance as a system, are dependent on the mobility patterns of the host vehicles \citep{ADRMdR2018}. Understandably, the movement of  third-party vehicles (taxis, buses, logistic vehicles) are confined to their designated routes or service-oriented operations, which are not specifically aligned with sensing objectives. Therefore, the allocation of sensors to such vehicles (strategic), as well as their full or partial maneuvers (tactical or operational), constitute an important aspect of vehicle-based urban sensing. Many such issues can be addressed as optimization problems, with varying degrees of spatial-temporal granularity, decision resolution, information availability, and user preferences. The following four questions are most commonly addressed in the literature, which are essential for the development and adoption of drive-by sensing systems:
\begin{itemize}
\item[Q1.] How to quantify the spatial-temporal sensing quality? 
\item[Q2.] What is the sensing power of a vehicle fleet of certain type (e.g. taxi, bus, UAVs)?
\item[Q3.] How to deploy/allocate a set of sensors to a vehicle fleet?
\item[Q4.] How to operationally maneuver a vehicle fleet to increase their sensing power? 
\end{itemize}

\begin{figure}[h!]
\centering
\includegraphics[width=0.95\textwidth]{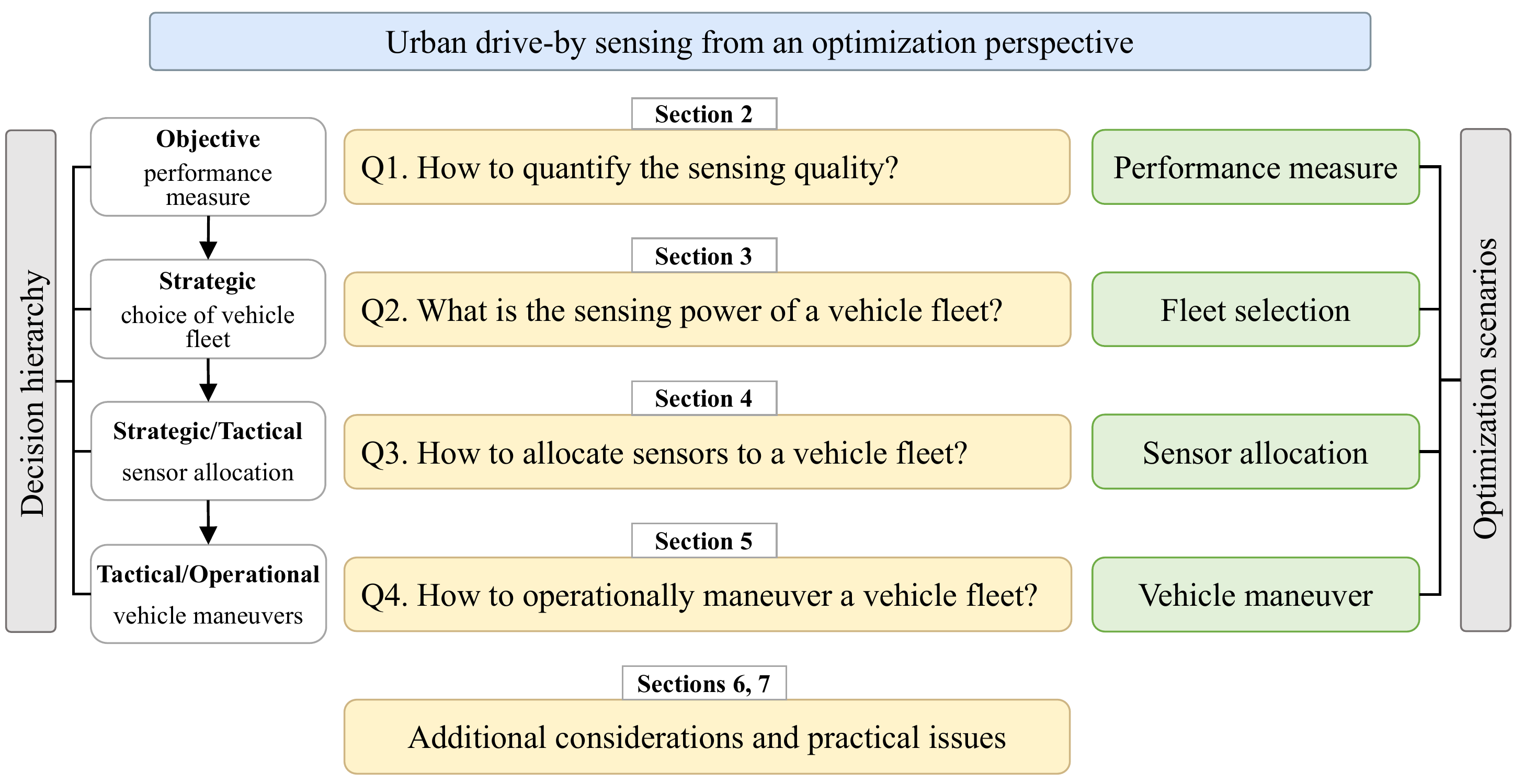}
\caption{Organization of main contents}
\label{figflow}
\end{figure}

The logical flow of these four questions is illustrated in Figure \ref{figflow}. Q1 serves as the foundation of optimization approaches, as various sensing preferences or objectives lead to different deployment and maneuver strategies and could even influence the choice of sensing vehicles. Q2 offers insights into DS by profiling different types of vehicle fleet in terms of the extent and granularity of their spatial-temporal coverage, operational reliability, cost-effectiveness and flexibility. Q3 is most commonly investigated in the literature, especially for taxis and buses without operational intervention. Q4 concerns with full or partial control over the fleet (e.g. vehicle routing and  scheduling, incentivizing schemes) in the interest of sensing quality improvement.

These four questions also reflect a natural order of decision making, starting from the performance measure (Q1-optimization objective), then the choice of vehicle fleet (Q2-strategic), followed by the sensor allocation (Q3-strategic/tactical), and furthered with vehicle maneuvers (Q4-tactical/operational). For a given urban sensing application with certain budget, the optimal or ideal distribution of the sensors in a space-time domain is usually difficult to achieve in reality, due to the physical constraints imposed by the mobility patterns of the host vehicles. Therefore, the essential purpose of optimization, whether regarding fleet selection (Q2), sensor allocation (Q3), or vehicle maneuver (Q4), is to remove those constraints as much as possible, such that the resulting sensing efficacy is engineered towards its optimal state. This motivates this paper to take on an optimization perspective in portraying drive-by sensing.

We offer the following discussion in the remainder of this paper. Section \ref{secQuant} reviews quantitative approaches that assess the goodness of sensing (Q1). Section \ref{secpower} presents results and insights regarding the sensing capabilities of different types of vehicle fleets (Q2). Section \ref{secdeploy} focuses on the optimization problem of allocating sensors to a chosen fleet (Q3). Section \ref{secinterv} discusses measures and mechanisms that influence the operation of sensing fleets (Q4). Section \ref{secissue} offer some further discussion on theoretical and practical issues in drive-by sensing, following by concluding remarks in Section \ref{secconclu}

\section{Quantifying drive-by sensing quality}\label{secQuant}

Quantification of the sensing quality is a prerequisite of DS system deployment. The data requirement of drive-by sensing depends on the subject of sensing and the underlying application \citep{Restuccia2017}. For example, in crowdsourced traffic state (speed, travel time) monitoring, typically based on fine-grained GPS trajectory data, it is critical to collect samples that cover most main streets at a relatively high frequency since traffic states changes dynamically. In air quality or urban heat island sensing, the requirement on spatial-temporal resolution of the data is not as high, since pollutant concentration or temperature tend to vary relatively smoothly across the spatial-time domain.

In this section, we provide a unified quantification framework that subsume different approaches in the drive-by sensing literature. The area of interest is partitioned into subregions $g\in G$ and the analysis horizon is discretized into intervals $t\in T$. In the literature, $g$ can be treated as a square region (i.e. mesh, grid) \citep{C2020, ZMLL2015} or a street segment \citep{OASSR2019, AD2017}. In the following subsections, we present three types of quantification methods, based on the sample size, temporal gap, and measurement distributions, respectively. We also use the term `sensing score' to refer to the goodness of sensing in a general setting.

\subsection{Sensing score as a function of sample size}

The first method quantifies the sensing quality as a non-decreasing function of the number of visits (samples) each sub-region receives. Specifically, let $N_{gt}$ be the number of sensing vehicles covering region $g$ during time interval $t$. The corresponding sensing score, expressed as a function $\xi(\cdot)$, reads: 
\begin{equation}\label{eqnxi}
\xi\big(N_{gt}\big)=
\begin{cases}
1 \qquad &\hbox{if}~N_{gt}\geq K
\\
0\qquad & \hbox{otherwise}
\end{cases}
\end{equation}

\noindent Eqn \eqref{eqnxi} uses a cut-off threshold $K$ to determine 0-1 sensing scores. In many studies \citep{AD2017,TRMSO2020} $K$ is set to be 1, so that a single visit achieves the maximum score. In this case, the sensing objective focuses almost entirely on spatial extent of the coverage. Other studies \citep{Cruz2020a, ZML2014} consider $K >1$, which is in the interest of more frequent and reliable sampling achieved by multiple visits. As $K$ increases, the focus shifts from spatial coverage to sample size per unit area, or the temporal frequency of data collection.

The overall spatial-temporal coverage  $\phi$ can be quantified as \citep{DH2023}
\begin{equation}\label{eqnss}
\phi=\sum_{t\in T}\mu_t \sum_{g\in G}w_g \xi\big(N_{gt}\big)
\end{equation}
\noindent where $\mu_t$ and $w_g$ are, respectively, temporal and spatial weights assigned to the discrete units in the space-time domain. These weights should be determined jointly by the sensing application scenario (e.g. air quality or traffic state monitoring) and the specific situation of the target area.

\subsection{Sensing score as a function of temporal gap}

Another line of research focus on the temporal distribution of sensing measurements, by considering the so-called {\it gap interval}, during which a sub-region $g\in G$ is not covered by any sensors. Specifically, consider a time interval $t\in T$, during which $g$ is covered by $m$ sensing vehicles. Let $s_i$ be the length of the i-th gap interval, and  $\mathcal{T}_{g,t} =\{s_1,\ldots,  s_n\}$ be the set of gap interval lengths during $t\in T$. Then, the sensing quality can be quantified in terms of the following quantities:
\begin{itemize}
\item The maximum gap between two consecutive visits \citep{PDB2012}: $\hbox{MG}_{g,t}=\max_{1\leq i\leq n} \{s_1,\ldots,s_n\}$, where smaller $\hbox{MG}_{g,t}$ implies higher sensing quality;

\item The gap averaged over the number of vehicles \citep{ADFS2021,Mathur2010}: $\hbox{AG}_{g,t}=\sum_{i=1}^n s_i / m$, where smaller $\hbox{AG}_{g,t}$ indicates higher sensing quality;

\item The averaged gap with buffers \citep{J2021}: $\hbox{AGB}_{g,t}=\sum_{i=1}^n\max\{0,\,s_i-b\}/m$, where $b$ is a fixed buffer such that any gap interval shorter than $b$ is ignored from the calculation (the buffer $b$ can be tailored to depend on $t$ or $g$ to diversify the sensing requirements);

\item The empirical distribution (cumulative distribution function) of the gap interval lengths \citep{ZMLL2015}: $F_{g,t}(\tau;\,m)=P\big\{s_i \leq \tau\vert \forall 1\leq i\leq n \big\}$.
\end{itemize}

For the first three quantification methods, the overall sensing score can be calculated as 
\begin{equation}
\phi=\sum_{t\in T}\mu_t \sum_{g\in G}w_g\hbox{MG}_{g,t}\quad \hbox{or}\quad \sum_{t\in T}\mu_t \sum_{g\in G}w_g\hbox{AG}_{g,t}\quad \hbox{or}\quad \sum_{t\in T}\mu_t \sum_{g\in G}w_g\hbox{AGB}_{g,t}
\end{equation}

\subsection{Sensing score measured as the difference from target distribution}

The third method differs considerably from the first two by focusing on the spatial-temporal distribution of the sensing scores. This approach typically requires knowledge of an a priori (target) distribution, and some form of dissimilarity measure of two distributions. For example, \citep{XCPJZN2019} consider the Kullback-Leibler (KL) divergence between the distribution of actual sensing rewards and a target distribution. The KL-divergence from Bayesian inference \citep{KL1951} measures the change of information when one revises beliefs from the prior distribution $O$ to the posterior distribution $P$. 

Formally, recalling that $N_{gt}$ denotes the number of visits of region $g$ during time interval $t$, we let $\zeta(\cdot)$ be an arbitrary sensing reward function. Then define the distribution
\begin{equation}
P(g,t)={ \zeta\big(N_{gt}\big)\over \sum_{g'\in G}\sum_{t'\in T} \zeta\big(N_{g't'}\big)}\qquad\forall g\in G,\,t\in T
\end{equation}
\noindent Let $O(g,t)$ be a discrete target distribution defined on the domain $G \times T$. The KL divergence of $P$ from $O$ is
\begin{equation}\label{eqnKL}
\hbox{KL}(P|| O)=\sum_{g\in G}\sum_{t\in T}P(g,t)\log {P(g,t)\over O(g,t)}
\end{equation}

\subsection{Discussion}

The selection of quantification measures for sensing quality is closely related to the underlying sensing application. For example, traffic congestion and air quality monitoring require repeated sampling of the same sub-region (or road segment) since each sample (vehicle visit) may be subject to unobserved randomness. Therefore, the sensing quality can be quantized using the first method \eqref{eqnxi}. As another example, quantities such as parking availability require high-frequency monitoring, and the second method based on the temporal spacing of visits would be appropriate. The third method is applicable when a target distribution of sensing resources is given, which typically arises in comprehensive area survey such as air quality and urban heat island monitoring. However, we note that, diversified sensing priorities or requirements even for the same sensing subject may lead to different evaluation methods. Table \ref{tabsqq} categorizes relevant literature by their sensing evaluation methods, vehicles used and sensing application. 

\begin{table}[h!]
		\centering
		\caption{Different sensing quality quantification methods adopted in the literature.}
		\small{
		\begin{tabular}{|m{0.29\textwidth}| m{0.14\textwidth}<{\centering} |m{0.11\textwidth}<{\centering} |m{0.33\textwidth}<{\centering}|}
		\hline
		Study & Quantification methods & Fleet type & Sensing scenario
		\\
		\hline
		\makecell[l]{\cite{YFLRL2021}} & sample size & Taxi & Traffic condition
		\\
		\hline
		\makecell[l]{\cite{GQDRJ2022}} & sample size & Taxi & Traffic condition
            \\
		\hline
		\makecell[l]{\cite{Singh2020}} & sample size & Taxi & Not specified
		\\
		\hline
		\makecell[l]{\cite{WLQWL2018}} & sample size & Bus & Traffic condition
		\\
		\hline
		\makecell[l]{\cite{Gao2016}} & sample size & Bus & Air quality
		\\
		\hline
		\makecell[l]{\cite{YLL2012}} & sample size & Bus & Air quality
		\\
		\hline
		\makecell[l]{\cite{Kaivonen2020}} & sample size & Bus & Air quality
		\\
		\hline
		\makecell[l]{\cite{Agarwal2020}} & sample size & Taxi \& bus & Air quality
		\\
		\hline
		\makecell[l]{\cite{CDMFD2018}} & sample size & Bus & Air quality; traffic condition
		\\
		\hline
		\makecell[l]{\cite{Cruz2020a}} & sample size & Bus & Waste disposal; air quality; noise
		\\
		\hline
		\makecell[l]{\cite{Cruz2020b}} & sample size & Bus & Waste disposal; Air quality; noise 
		\\
		\hline
		\makecell[l]{\cite{AD2017}} & sample size & Bus & Road surface
		\\
		\hline
		\makecell[l]{\cite{TRMSO2020}} & sample size & Bus & Heat island
		\\
		\hline
		\makecell[l]{\cite{Trotta2018}} & sample size & UAV & Built environment
		\\
		\hline
		\makecell[l]{\cite{Rashid2020}} & sample size & UAV & Disaster response
		\\
		\hline
		\makecell[l]{\cite{Mathur2010}} & temporal gap & Taxi & Parking availability
		\\
		\hline
		\makecell[l]{\cite{BAD2017}} & temporal gap & Taxi & Parking availability
		\\
  	\hline
		\makecell[l]{\cite{PDB2012}} & temporal gap & Robot & Built environment
		\\
  	\hline
		\makecell[l]{\cite{ADFS2021}} & temporal gap & Taxi & Not specified
		\\
  	\hline
	\makecell[l]{\cite{J2021}} & temporal gap & Dedicated vehicle & Air quality
		\\
  	\hline
		\makecell[l]{\cite{ZMLL2015}} & temporal gap & Taxi & Not specified
		\\
  	\hline
		\makecell[l]{\cite{XCPJZN2019}} & target distribution & Taxi & Air quality; built environment
		\\
		\hline
		\end{tabular}
		}
		\label{tabsqq}
\end{table}

\section{Assessing the sensing power of vehicle fleets}\label{secpower}

In urban drive-by sensing, the selection of vehicle fleet to carry out the sensing task is a strategic decision, which relies on knowledge of the candidates' sensing capabilities, as well as the set-up and maintenance costs of instrumentation, among other case-specific considerations. \cite{ADRMdR2018} discuss the applicability of various third-party vehicles (taxis, buses, and service vehicles) in ubiquitous sensing,  and find that the mobility patterns of host vehicles play a major role in the efficacy of drive-by sensing. This section reviews some empirical studies exploring the sensing power of public transport (taxi and bus) or for-hire (e.g. UAVs) vehicles.

\subsection{Taxi fleets}

Taxis have become the most widely used host vehicles for their high spatial mobility, long operational hours and low deployment and maintenance costs. Some experimental studies have evaluated the sensing power of taxi fleets in cities by analyzing actual taxi trajectories. \cite{BBLARB2016} use real-world taxi trajectories over a six-month period in the city of Rome, and show that even a relative small fleet (120 taxis) can achieve 80\% coverage of the downtown area in 24 hours. \cite{BAD2017} explored the spatial-temporal movement pattern of taxis in San Francisco and evaluated their suitability for sensing on-street parking availability, which reveal a strong dependence of taxi coverage on the hour of day and road class. In addition, despite their heterogeneous space-time distribution, about 500 taxis can adequately cover about 90\% of road segments. \cite{Martino2022} analyze trajectories of taxi fleets in the city of Porto, and show that just 100 taxis can achieve a remarkable spatial coverage during certain periods, but the temporal coverage falls short for some urban sensing tasks. \cite{ZMLL2015} define the opportunity coverage ratio as the expected proportion of grid cells that can be covered in each time interval. Then, they estimate the required number of taxis in Beijing and Shanghai to be 1700 and 1900, respectively, to achieve an opportunity coverage ratio of no less than 50\% in an hour. \cite{OASSR2019} propose a random search process of taxis and a ball-in-bin model to quantify the sensing power of taxi fleets. By analyzing empirical taxi trajectory data from 9 cities around the globe, the authors draw the following conclusion: (1) Only a few taxis can achieve a high sensing coverage, but the number of taxis required increase sharply as the monitoring range and frequency increase. (2) taxi fleets have large but limited sensing power: popular street segments are easily covered, but unpopular segments are rarely surveyed, and some may even be never scanned.

\subsection{Bus fleets}
	
	Buses are deemed suitable hosts for drive-by ubiquitous sensing because of the extensive spatial coverage of bus routes, reliable bus operations (based on timetables), and highly predictable bus trajectories. Similar to taxis, actual bus trajectories have been used to analyze the sensing power of bus fleets. \cite{CDMFD2018} analyze the spatial-temporal coverage of bus fleet in Rio de Janeiro in Brazil, and found that (1) the entire fleet covers approximately 47\% of the urban streets; (2) 18\% of the fleet contribute to 94\% of the total coverage; (3) beyond certain fleet size, the coverage tends to concentrate on central areas instead of the entire region. \cite{Cruz2020b} quantify the contribution of buses to drive-by sensing under different applications (waste management, air quality, and noise monitoring) in Rio de Janeiro, Brazil, and the results indicated that the contributions of individual buses vary greatly across these application scenarios. \cite{Cruz2020a} propose a coverage metric that simultaneously considers the transmission delays and sensing frequency, which is used alongside bus trajectory data in Rio de Janeiro to quantify the sensing quality in different application scenarios. The authors demonstrate the considerable impact of sensing requirements specific to each application, while showing that bus-based sensing has distinctive advantage over fixed-location (stationary) sensing for some urban applications. \cite{JLH2022} explore the sensing power of bus fleets in central Chengdu, China, by adopting a sequential optimization approach with three sub-problems: (1) bus route selection, (2) minimum fleet size, and (3) sensor allocation. The authors partition the area into 219 spatial grids of size 1km$\times$1km, and show that only 30-40 sensors are sufficient to cover every grid at least once in an hour.

\subsection{UAVs and DVs}

In some urban sensing applications, it is crucial to survey highly specific target regions during certain times, instead of pervasive sensing. One example is urban built environment sensing for unauthorized construction and waste disposal activities. Another example involves noise or air quality monitoring at sensitive sites such as factories. Such sensing tasks are usually performed by dedicated vehicles (DVs) or unmanned aerial vehicles (UAVs) for their high controllability and flexibility in performing targeted sensing.

\cite{YZBSH2018} design a mobile air quality index (AQI) monitoring system based on UAVs to construct fine-grained AQI maps. The experimental results showed that the system can provide higher accuracy in AQI mapping than fixed-location sensing networks. \cite{HBYZBS2019} present an aerial-ground air quality sensing system based on both fixed stations and UAVs. The experimental results show that: (1) Aerial sensing can well complement ground sensing to provide fine-grained monitoring; (2) The efficacy of aerial sensing is primarily limited by battery capacity of UAVs. \cite{Rashid2020} present an integrated system for reliable disaster response based on social media and UAVs sensing. Their main idea is to use UAVs to verify uncertain information extracted through social media. An instance of real-world disaster response application showed that this system significantly outperformed both the social-only and UAV-only solutions. For DVs, \cite{FZGJGW2021} use a taxi-DV hybrid sensing method for fine-grained spatial-temporal sensing. The experimental results using a real-word dataset from Shenzhen, China showed that DVs can complement spatially unbalanced sensing offered by taxi fleets.

\subsection{Discussion}

A broad notion of sensing efficacy can be characterized in various aspects by the following 6 indicators
\begin{enumerate}
\item Spatial extent: The size of area that can be frequently scanned by the fleet. 
\item Temporal duration: The maximum time span during which an area can be continuously scanned by a fleet.
\item Sensing reliability: The likelihood that a given sub-region within the reach of the fleet can be scanned at least once within certain time interval. 
\item Spatial resolution: The smallest spatial unit that can be scanned at least once. 
\item Sensing flexibility: The maneuverability of the fleet to scan a target point in space and time.
\item Cost effectiveness: Purchase, operational and maintenance costs (resulting from the sensing functionality only) to collect a unit amount of data. 
\end{enumerate}

Table \ref{tabPI} summarizes relevant data and conclusion found in the literature, which can be used to partially indicate the performance of different fleet types (very low, low, medium, high, very high). 

\small{
\begin{longtable}{m{0.12\textwidth}| m{0.05\textwidth}<{\centering} |m{0.12\textwidth}<{\centering} |m{0.15\textwidth}<{\centering} |m{0.43\textwidth}<{}}
\caption{Relevant data and conclusion from the literature on different performance indicators of various fleet types.}
\label{tabPI}
\\
        \hline
        Study	& Fleet type & Sensing scenarios &Performance indicator & Description  \\
        \hline
        \endhead
        \cite{ADRMdR2018} & Taxi & Air quality; Road surface & Spat. res. (high) & In Manhattan, only five taxis can cover around 30\% of street segments at least once per day. With 30 taxis, such a number increases to around 60\%.
    	\\
    	\hline
    	\cite{BBLARB2016} & Taxi & Built environment & Spat. ext. (medium) &  In Rome, 120 taxis can achieve 80\% coverage of the downtown area in 24 hours. 
    	\\
    	\hline
    	\cite{BAD2017} & Taxi & Parking availability & Spat. res. (high) &  In San Francisco, about 500 taxis can adequately cover about 90\% of road segments. 
    	\\
    	\hline
    	\cite{Martino2022} & Taxi & General & Sens. rel. (medium) &   In Porto, just 100 taxis can achieve a remarkable spatial coverage during certain periods, but the temporal coverage falls short for some urban sensing tasks.
    	\\
    	\hline
    	\cite{ZMLL2015} & Taxi & General & Sens. rel. (medium) &  To achieve an opportunity coverage ratio of no less than 50\% in an hour, only 1700 and 1900 taxis are required in Beijing and Shanghai, respectively.
    	\\
    	\hline
    	\cite{OASSR2019} & Taxi & General & Spat. ext. (medium), sens. rel. (low) &  Taxi fleets have large but limited sensing power: popular street segments are easily covered, but unpopular segments are rarely surveyed, and some may even be never scanned.
    	\\
    	\hline
    	\cite{ADRMdR2018} & Bus & Air quality; road surface & Spat. flex. (low), sens. rel. (medium) & Buses cover their predefined routes, which consist of a fixed number of street segments, many times per day.
          \\
    	\hline
    	\cite{CDMFD2018} & Bus & Air quality; traffic condition & Spat. res. (medium); sens. flex. (low) & In Rio de Janeiro, (1) the entire fleet covers approximately 47\% of the urban streets; (2) 18\% of the fleet contribute to 94\% of the total coverage.
          \\
    	\hline
    	\cite{Cruz2020b} & Bus & Waste disposal; air quality; noise & Sens. rel. (medium) & In Rio de Janeiro, the contributions of individual buses vary greatly across these application scenarios.
          \\
    	\hline
    	\cite{Cruz2020a} & Bus & Waste disposal; air quality; noise & Spat. ext. (high) & Bus-based sensing increases the coverage by up to 7.6 times compared to fixed-location (stationary) sensing.
          \\
    	\hline
    	\cite{JLH2022} & Bus & General & Sens. rel. (high) & 
	Only 30-40 instrumented buses are sufficient to cover every one of 219 spatial grids (1km$\times$1km) at least once in an hour.
          \\
    	\hline
    	\cite{YZBSH2018} & UAV & Air quality & Spat. res. (very high), sens. flex. (very high) & The AQI monitoring system based on UAVs can provide finer-grained AQI maps than fixed-location sensing networks.
          \\
    	\hline
    	\cite{HBYZBS2019} & UAV & Air quality & Spat. res. (very high), sens. flex. (very high) & Aerial sensing can well complement ground sensing to provide fine-grained monitoring.
          \\
    	\hline
    	\cite{FZGJGW2021} & DV & Air quality, traffic condition & Sens. flex. (very high) & DVs can complement spatially unbalanced sensing offered by taxi fleets. 
          \\
    	\hline
    	\cite{JHG2023} & DV & General & Sens. flex. (very high) & DVs can be flexibly navigated to monitor a set of points of interest with arbitrary sensing weights.
          \\
    	\hline
    	\cite{ADRMdR2018} & Trash truck & Air quality; road surface & Temp. dur. (low) & Trash trucks cover a larger number of street segments but operate for fewer hours per day, and usually operate only a few days a week in each zone.
        \\
        \hline
    \end{longtable}
    }

\begin{figure}[h!]
\centering
\includegraphics[width=.7\textwidth]{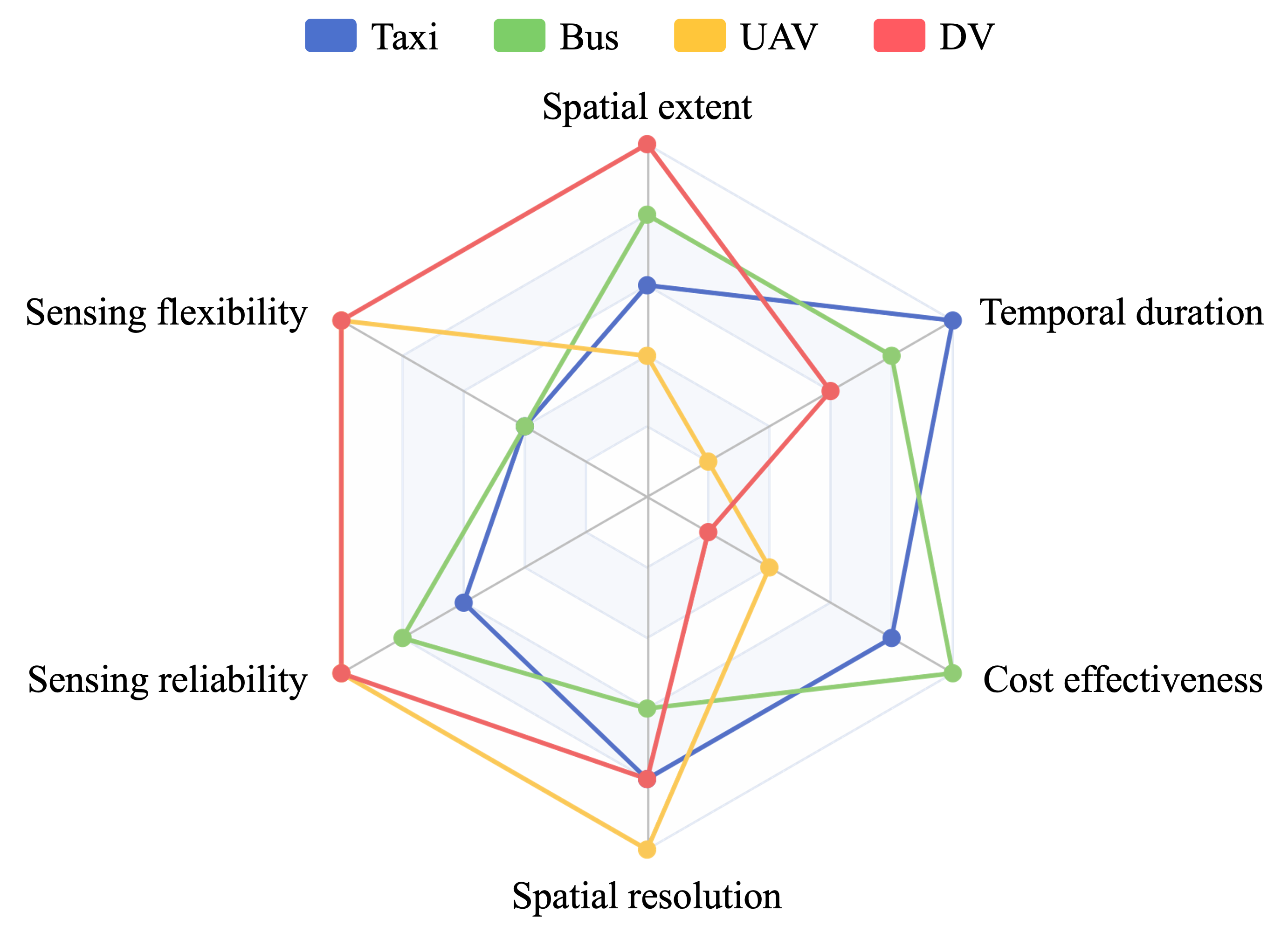}
\caption{Comparison of different fleet types in drive-by sensing.}
\label{figradar}
\end{figure}

Building on Table \ref{tabPI}, together with some intuition and practical experience, in Figure \ref{figradar} we qualitatively compare bus, taxi, unmanned aerial vehicles (UAVs) and dedicated vehicles (DVs) in terms of the 6 indicators. In particular, 
\begin{itemize}
\item Bus fleets excel in spatial extent, temporal duration and cost effectiveness, as buses have routes that extend to remote areas with fixed timetables that boost their temporal duration and sensing reliability.
\item Taxi fleets, in addition to cost effectiveness like buses, enjoy superior temporal duration and spatial resolution, as they have very long operating hours as well as flexible routes that cover most urban streets. 

\item When it comes to costs, both buses and taxis do not require much investment beyond initial installation and regular maintenance. However, in our practical experience, the installation and maintenance fees for taxis are slightly higher because taxis are more revenue-driven and charge more for the time/opportunity lost due to service disruptions.

\item Both UAVs and DVs have high sensing flexibility, reliability and resolution, since they are fully controllable to fulfill very specific sensing requirements. However, they are quite costly and require additional human work to control them. Moreover, the mobility of UAVs is typically constrained by limited battery capacity, resulting in low spatial extent and temporal duration. 
\end{itemize}

We consider several applications of drive-by sensing found in the literature, including air quality \citep{SHS2021}, traffic condition \citep{GQDRJ2022}, heat island phenomena \citep{FBRDRAP2021}, road surface conditions \citep{Eriksson2008}, natural/built environment \citep{Rashid2020} and on-street parking availability \citep{Mathur2010}. Table \ref{tabapp} outlines the relative importance of the 6 attributes for each application scenario. Note that the contents of Table \ref{tabapp} are qualitative and for illustration purposes only. Combining Figure \ref{figradar} with such relative importance renders recommendation for the type of host vehicles, as shown in the last column of Table \ref{tabapp}.

\begin{table}[h]
\centering
\caption{Relative importance (indicated by the number of check marks) of the 6 attributes in several drive-by sensing scenarios.}
\begin{tabular}{ccccccc|c}
\hline
Sensing  & Spat. & Sens. & Sens. & Spat.& Cost & Temp. & Suitable
\\
scenarios & extent & flex. & rel. & res.& eff. & dur. &  fleet
\\
\hline
Air quality & \checkmark\checkmark\checkmark &  &  &  & \checkmark\checkmark & \checkmark\checkmark\checkmark & Bus, taxi
\\
\hline
Traffic condition & \checkmark\checkmark & & \checkmark & \checkmark\checkmark\checkmark & \checkmark\checkmark\checkmark & \checkmark\checkmark\checkmark & Taxi
\\
\hline
Heat island & \checkmark\checkmark\checkmark & \checkmark& & & \checkmark\checkmark & \checkmark\checkmark & Bus, taxi
\\
\hline
Road surface & \checkmark\checkmark\checkmark & \checkmark & & \checkmark\checkmark\checkmark & \checkmark & & DV
\\
\hline
Built environment & \checkmark & \checkmark\checkmark\checkmark & & \checkmark\checkmark\checkmark & & & UAV
\\
\hline
Parking availability & \checkmark\checkmark & \checkmark & \checkmark\checkmark\checkmark & \checkmark\checkmark & \checkmark\checkmark & \checkmark\checkmark & Mixture
\\\hline
\end{tabular}
\label{tabapp}
\end{table}

\section{Sensor deployment problems in drive-by sensing}\label{secdeploy}

As another instance of strategic decision making in drive-by sensing, the sensor deployment problems are widely investigated in the literature. They can be interpreted as non-trivial extensions of the classical facility location problems in the logistics literature \citep{WWWZQ2021}. This section starts with a conceptual yet general form of the sensor deployment problem, followed by discussion of its applications in sensor allocation for taxis and transit vehicles.

\subsection{A conceptual model of sensor deployment}

The sensor deployment problem aims to allocate a fixed number of sensors (or budget) to a fleet of vehicles, such that the overall sensing quality is optimized. Two critical components of this problem are: 
\begin{itemize}
\item The mathematical articulation of sensing quality, which is expressed as a function (functional) of the sensors' trajectories in space and time (see Section \ref{secQuant} for more details);
\item The mobility patterns of the host vehicles, which differ drastically by vehicle type (i.e. taxi, bus, dedicated vehicles, UAVs), and their mathematical expressions as constraints.
\end{itemize}

We consider a generic form of the sensor deployment problem: assigns $N$ sensor to $M$ candidate vehicles such that the overall sensing quality $\phi$ is maximized (of course, we assume that each vehicle can be equipped with at most one sensor and, in most cases, $M\gg N$). In the following conceptual model, the decision variables $x_i$ equals 1 if vehicle $i$ is equipped with sensor and 0 otherwise. Eqn. \eqref{conm1} expresses the sensing quality maximization based on $N_g$, which denotes vehicle coverage at sub-region $g\in G$. In \eqref{conm2}, $N_g$ is the sum of coverage by individual sensing vehicles $N_g^i$. Eqn \eqref{conm3} expresses $N_g^i$ using a function ${\bf M}_g$ that encapsulates the mobility pattern of the $i$-th vehicle. For example, if the candidate vehicles are taxis, then the coverage of sub-region $g$ by vehicle $i$ can be expressed using a probability model following a ball-in-bin approach \citep{OASSR2019}; for buses, $N_g^i$ can be determined based on the route information of bus $i$. The sensor limit constraint is expressed in \eqref{conm4} and the binary nature of the decision variables are stated in \eqref{conm5}.
\begin{align}
\label{conm1}
& \max_{{\bf x}=(x_1,\ldots, x_M)} \phi\big(N_g: g\in G \big)
\\
\label{conm2}
& N_g = \sum_{i=1}^M N_g^i \quad \forall g\in G
\\
\label{conm3}
& N_{g}^i={\bf M}_g (x_i)\quad \forall g\in G,~i=1,\ldots,M
\\
\label{conm4}
& \sum_{i=1}^M x_i \leq N
\\
\label{conm5}
& x_i \in \{0,\,1\}\quad\forall i=1,\ldots, M
\end{align}

\begin{remark}
The binary decision variable $x_i$'s are not necessarily associated with individual vehicles. For example,  studies on bus-based drive-by sensing \citep{YLL2012} consider $M$ bus lines and let $x_i=1$ if the $i$-th line is instrumented with sensors. In this case, the mobility equation $N_g^i={\bf M}_g(x_i)$ reduces to the following incidence variables:
$$
N_g^i=\delta_{gi}=
\begin{cases}
1\qquad \hbox{if Line } i \hbox{ covers region }g
\\
0\qquad\hbox{otherwise}
\end{cases}
$$
The corresponding optimization problem is also known as the set covering problem \citep{Beasley1996,Caprara1999}. 
 \end{remark}
 
 The conceptual model \eqref{conm1}-\eqref{conm5} serves as a basic starting point of many sensor deployment problems investigated in the literature, which will be elaborated in the following sections.

\subsection{Sensor deployment problems for taxis}

The sensor deployment problem for taxis is typically formulated as a subset selection problem. The premise is that, while taxi trajectories are in general difficult to predict, the mobility patterns of individual taxis might be heterogeneous, depending on the driver's personal preferences and other unobserved factors. Therefore, by analyzing historical taxi trajectories, one may select a subset of taxis that promise good sensing outcome. 

\cite{ZML2014} propose a vehicle selection mechanism, which determines the minimum number of taxis to achieve coverage requirements. \cite{HCL2015} propose a greedy approximation algorithm and a genetic algorithm to solve the participant selection problem with a fixed budget to achieve the optimal coverage based on predicted trajectory. \cite{KGB2016} propose an information-centric algorithm to measure the taxis’ relative importance in the network, then presented a selection algorithm to find the best subset of taxis to achieve a desired coverage within a fixed budget. \cite{LNL2016} propose a vehicle selection method that uses heterogeneous vehicle mobility information forecasted by a continuous-time Markov chain for the collection of comprehensive tempo-spatial sensing data. They consider the vehicle selection issue as a knapsack problem and solve it with a polynomial-time greedy approximation algorithm. \cite{YDLCG2017} designed a greedy approximation algorithm with linear-time complexity to select taxis by minimizing costs while optimizing spatial-temporal coverage. \cite{ZYL2018} modeled the vehicle selection problem as a bi-objective approach, which jointly optimizes the total coverage of all vehicles and the reliability of individual sensing tasks (considering possible non-compliance on part of the drivers). \cite{YFLRL2021} consider traffic condition monitoring and propose a vehicle selection model based on the spatial-temporal characteristics of the road network, which takes into account the speeds of the host vehicles and their impact on the sensing quality. \cite{CLGZX2017} postulate a scenario where the sensor can be manually switched on and off, and the sensing cost is proportional to the length of the vehicle trajectory during which the sensor is on. Then, they propose a greedy heuristic to select appropriate trajectory segments to maximize the sensing quality at limited costs.

\subsection{Sensor deployment problems for transit vehicles} 

For overground transit vehicles (e.g. buses, trams), the sensor deployment problem mainly focuses on allocating sensors to transit lines or individual vehicles. Depending on the decision resolution, these studies can be categorized as line selection and vehicle selection. 

Regarding line selection, \cite{YLL2012} use a chemical reaction optimization approach to select a subset of bus lines to install sensors. \cite{AD2017} design a greedy heuristic algorithm to install a limited number of sensors on the bus lines to maximize the total number of segments in road surface condition monitoring. \cite{Kaivonen2020} use an image analysis algorithm to evaluate the percentage of area that can be covered by different bus route combinations. \cite{Saukh2012} designed an evolutionary algorithm to select a subset of tram lines based on the service timetables to achieve good coverage with data calibration constraints (see Section \ref{subsecDatacal}).

Regarding vehicle selection, \cite{Gao2016} and \cite{WLQWL2018} design a greedy and an approximate algorithms, respectively, to select a subset of buses to maximize the coverage utility. \cite{TRMSO2020} design a heuristic algorithm to allocate a limited number of sensors to buses considering the importance level of different geographical regions in monitoring the urban heat island phenomenon. \cite{Agarwal2020} aim to select a subset of vehicles from a hybrid fleet of buses and taxis to maximize the spatial-temporal coverage of the city. They formulated this problem as an integer linear programming model and designed an approximate algorithm to solve it.

Different from the aforementioned models, \cite{JHL2023} adopt a trip-based approach, which explicitly consider timetabled trips that must be fulfilled by the bus fleet while a portion of them perform sensing tasks. To address the computational challenge in large-scale instances, the authors design a multi-stage solution framework that decouples the spatial-temporal structures of the sensing task through line pre-selection and bi-level optimization. As a result, the computational complexity is reduced to be sub-linear w.r.t. the number of lines, rather than combinatorial w.r.t. the number of buses in existing vehicle selection approaches.

\subsection{Synopsis}

Table \ref{tabsdp} summarizes relevant literature on the sensor location problem for various vehicle types. They all focused on allocating sensors to vehicles or bus/tram lines based on actual or predicted vehicle trajectories and employ heuristic, meta-heuristic, or approximate algorithms to solve the problem. It is noted that all these studies fall within the category of opportunistic sensing \citep{MZY2014, Lane2008}.

\begin{table}[h!]
		\centering
		\caption{Summary of sensor deployment problems.}
		\small{
		\begin{tabular}{|m{0.26\textwidth} | m{0.29\textwidth} |m{0.07\textwidth}| m{0.26\textwidth}|}
		\hline
		Study & Dataset & Sensor host & Solution approach
		\\
		\hline
		\makecell[l]{\cite{ZML2014}} & GPS trajectories of 10,357 taxis in Beijing, China & Taxi & Greedy heuristic algorithm
		\\
		\hline
		\makecell[l]{\cite{HCL2015}} & Taxi trajectory data in Cologne, Germany & Taxi & Greedy approximation algorithm and genetic algorithm
		\\
		\hline
		\makecell[l]{\cite{KGB2016}} & Taxi trajectory data in Cologne, Germany & Taxi & Greedy heuristic algorithm
		\\
		\hline
		\makecell[l]{\cite{LNL2016}} & GPS trajectories of 10,357 taxis in Beijing, China & Taxi & Greedy approximation algorithm
		\\
		\hline
		\makecell[l]{\cite{YDLCG2017}} & GPS trajectories of 4,316 taxis in Shanghai, China & Taxi & Greedy approximation algorithm
		\\
		\hline
		\makecell[l]{\cite{ZYL2018}} & GPS trajectories of 10,357 taxis in Beijing, China & Taxi & Greedy heuristic and multi-objective genetic algorithm (NSGA-II)
		\\
		\hline
		\makecell[l]{\cite{YFLRL2021}} & Taxi trajectories in Cologne, Germany & Taxi & Greedy heuristic algorithm
		\\
		\hline
		\makecell[l]{\cite{CLGZX2017}} & Taxi trajectories in Rome, Italy & Taxi & Greedy heuristic algorithm
		\\
		\hline
		\makecell[l]{\cite{YLL2012}} & 91 bus lines in Hong Kong Island, China & Bus line & Chemical Reaction Optimization (meta-heuristic)
		\\
		\hline
		\makecell[l]{\cite{AD2017}} & 713 bus lines in London, UK & Bus line & Greedy heuristic algorithm
		\\
		\hline
		\makecell[l]{\cite{Kaivonen2020}} & 21 bus lines in Uppsala, Sweden & Bus line & Image analysis algorithm
		\\
		\hline
		\makecell[l]{\cite{Gao2016}} & Trajectories of 1415 buses in Hangzhou, China & Bus & Greedy heuristic algorithm
		\\
		\hline
		\makecell[l]{\cite{WLQWL2018}} & Bus trajectories in Beijing, China & Bus & Greedy approximate algorithm
		\\
		\hline
		\makecell[l]{\cite{TRMSO2020}} & Trajectories of 20 buses in Athens, Georgia & Bus & Greedy heuristic algorithm
		\\
		\hline
		\makecell[l]{\cite{Saukh2012}} & 13 tram lines in Zurich, Switzerland & Tram line & Evolutionary algorithm
		\\
		\hline
		\makecell[l]{\cite{Agarwal2020}} & Taxi and bus trajectories in San Francisco, USA & Taxi \& bus & Approximation algorithm
		\\
		\hline
		\makecell[l]{\cite{JHL2023}} & Timetables of 167 bus lines in Chengdu, China & Bus trips & Multi-stage solution algorithm
		\\
		\hline
		\end{tabular}
		}
		\label{tabsdp}
\end{table}

\section{Operational intervention in drive-by sensing}\label{secinterv}

Instead of opportunistic sensing, which exerts no influence on the operations of the host vehicles, another line of research investigates the possibility of improving sensing quality by applying intervention measures, ranging from incentivizing schemes to active scheduling. The design of such interventions is highly dependent on the nature of the host fleet, i.e. institutional barriers, operational constraints, participant compliance, and cost-effectiveness.

\subsection{Intervention on taxi operations} 

Taxi-based drive-by sensing is associated with relatively low costs and high temporal duration (Figure \ref{figradar}), but its spatial survey is usually biased, primarily driven by factors such as spatially heterogeneous trip requests, and the tendency to follow shortest/popular paths. To improve the spatial coverage of taxi fleets while maintaining a satisfactory level of service, studies have resorted to incentivizing mechanisms and routing maneuvers.

\subsubsection{Incentivizing mechanism}
\cite{FJLQG2021} propose a system that integrates vehicle scheduling and incentive mechanisms to complete the sensing tasks with minimum cost. First, they formulate the sensing vehicle scheduling problem into an augmented set cover problem with spatial-temporal constraints by generating a driver's candidate trajectory set within an acceptable detour range when the origin-destination pair set is known. Then, they design a greedy algorithm to choose routes for vehicles and design a pricing algorithm based on reverse combinatorial auction to calculate the drivers’ rewards. \cite{C2020} focus on routing advice for vacant taxis with limited incentive budget, by proposing a greedy heuristic algorithm that selects appropriate subset of vacant taxis to incentivize, along with suggested routes, such that they contribute to the maximization of sensing quality with limited monetary incentives, which are estimated based on predicted deadhead trips and trip requests. A similar incentive mechanism is discussed in \cite{XCPJZN2019}, where the sensing quality maximization is replaced by minimization of the discrepancy between actual and target distribution of sensing rewards.

\subsubsection{Routing maneuvers}

One of the causes of unbalanced spatial survey afforded by taxi fleets is their tendency to follow shortest (or least costly) paths between origins and destinations, which often leads to overly concentrated coverage of certain main street segments, leaving others rarely visited. Studies have therefore investigated diversified routing choices following both decentralized and centralized decision architectures. \cite{ADFS2021} propose an $\varepsilon$-perturbed route set based on the A$^*$ algorithm, such that a trip with given origin and destination can be executed by following a route that contributes to greater sensing coverage without considerably deviating from the optimal routing choice. \cite{Masutani2015} consider a centralized decision framework where a traffic center collects information of planned routes of all relevant vehicles, and makes routing recommendations to all participants based on a sensing quality maximization procedure. Simulation results show that this method can harvest higher-coverage sensing data without extending vehicles' travel times. \cite{GQ2022} focus on the vacant cruising phase of taxi services, by devising a routing guidance framework for unoccupied taxis. The effectiveness of this method in boosting the sensing efficacy is demonstrated via a simulation study of ride-hailing fleet in New York.

\subsection{Intervention on bus operations}

Literature on the intervention of bus operations is sparse. \cite{DH2023} explore the sensing power of bus fleet by investigating the possibility of flexible sensor circulation in a transit network enabled by inter-line bus scheduling. They consider a fleet of normal and instrumented buses, and aim to maximize the sensing quality while minimizing bus operational costs (especially pertaining to inter-line relocations), while making sure that all timetabled trips are served at no cost of the service quality. This problem is formulated as a nonlinear integer program based on time-expanded network representation of the bus transit system \citep{LDLY2022}, which is solved by a batching scheduling heuristic algorithm. Using the bus network in Chengdu as a case study, the authors show that inter-line scheduling can improve the sensing quality by up to 32.1\% while slightly saving the operational cost. They further point out that sensor investment can be reduced by over 33\% when considering active bus scheduling. \cite{JHL2023} present a trip-based sensor allocation problem, which explicitly considers timetabled trips that must be executed by the fleet while a portion of them perform sensing tasks. To address the computational challenge in large-scale instances, we design a multi-stage solution frame- work that decouples the spatial-temporal structures of the sensing task through line pre- selection and bi-level optimization. As a result, the computational complexity is reduced to be sub-linear w.r.t. the number of lines, rather than combinatorial w.r.t. the number of buses in existing vehicle-based approaches. A real-world case study in central Chengdu demonstrates that coordinating bus scheduling and sensing tasks can substantially increase the spatial-temporal sensing coverage.

\subsection{Dedicated vehicles (DVs)} 
Dedicated sensing vehicles are suitable to (1) survey targeted areas for their high controllability, and (2) spatially and temporally complement the sensing of other fleet types. \cite{OASSR2019} proposed an idea of hybrid monitoring, where taxis are used to scan popular areas of a city, and hard-to-reach areas could be covered by dedicated vehicles. \cite{FZGJGW2021} further realized such an idea by proposing a repositioning policy for DVs to achieve fine-grained spatial-temporal sensing coverage at the minimum long-term operational cost. Considering the randomness of taxi trajectories, they adopt a robust optimization approach for the DV scheduling problem. \cite{TLH2021} design two real-time routing algorithms for a fleet of DVs based on multi-agent reinforcement learning. This approach is shown to significantly outperform heuristic random routing policies, and could be applied to a mixed fleet by complementing the coverage offered by taxis or buses, based on their real-time trajectory information. \cite{JHG2023} design an adaptive large neighborhood search algorithm and  a bilevel matheuristic method to solve the DV-based Volatile Organic Compounds (VOCs) monitoring, the results show that DVs can be flexibly navigated to monitor a set of points of interest with arbitrary sensing weights.

\subsection{Unmanned Aerial Vehicles (UAVs)}

The increased availability and readiness of general air mobility such as UAVs have raised their promises in urban sensing \citep{Lambey2021}. The advantages of UAV-based urban sensing include high controllability, which makes it suitable for targeted survey, and high mobility, which removes the physical constraint of street layouts.

 \cite{ZFGAMRG2018} consider a joint routing and area assignment problem in UAV-aided sensing, as a two-stage two-sided matching problem. The first stage divides the target area into sub-regions and design UAV routes using dynamic programming and genetic algorithm;  the second stage assigns UAVs to the routes.  \cite{XZMX2021} solve an urban sensing task allocation problem for UAVs, with heterogenous task requirements as input, using three heuristic algorithms  to minimize the sensing cost. \cite{DZCM2021} integrate community sensing with UAV-based sensing, and design an actor-centric heterogenous collaborative reinforcement learning algorithm to schedule different UAV sensors to maximize sensing coverage, coverage fairness and cost effectiveness. \cite{XZDQHCY2021} consider a scenario where UAVs are used for logistic delivery while simultaneously for urban sensing. The problem jointly determines package assignment, delivery routes and sensing windows to optimize delivery and sensing objectives, which is solved with an approximate algorithm. \cite{Trotta2018} propose a framework that allows UAVs to perform city-scale video monitoring of a set of Points of Interests (PoIs). They develop a mixed integer linear programming model to select the UAVs for periodic re-charging and repositioning by landing on buses.

\section{Relevant considerations for research and practice}\label{secissue}

\subsection{Drive-by sensing with spatial-temporal correlation of information}
Most studies that attempt to quantify the sensing quality using various metrics have not considered subsequent data-driven modeling or analyses that would benefit from such crowdsourced information. For example, in air quality sensing, mobile measurements of PM$_{2.5}$ concentrations were used alongside fixed-location measurements for spatial inference based on urban big data \citep{SHS2021}. In traffic state monitoring, information such as flow and speed on unvisited links is likely to be inferred (sometimes with remarkable accuracy) from data collected in neighboring links or nodes \citep{ZFZM2022,Viti2014,XLCC2016}. A few studies \citep{DZCM2021,GQ2022} have considered the effect of spatial and temporal correlation of measurements in mobile sensing, by postulating that measurements collected at certain point in space and time can be used to infer its neighbors. 

The significance of spatial or temporal inference, as undertaken in the aforementioned studies, is that it could change the way sensing coverage is evaluated. Integrating application-specific objectives or constraints would be an interesting and practical extension of urban drive-by sensing.

\subsection{Data calibration}\label{subsecDatacal}

In drive-by sensing, the accuracy and reliability of the data collected by mobile sensors is crucial for their intended application. To ensure satisfactory accuracy and consistency of the measurements, some mobile sensors need to be constantly calibrated by fixed reference stations via collocation \citep{Saukh2014, Maag2017, SHS2021}. This means that two sensors are placed in proximity in the space-time domain, and the measurement of one (the reference station) can be used to correct the other.

\begin{figure}[h!]
\centering
\includegraphics[width=.5\textwidth]{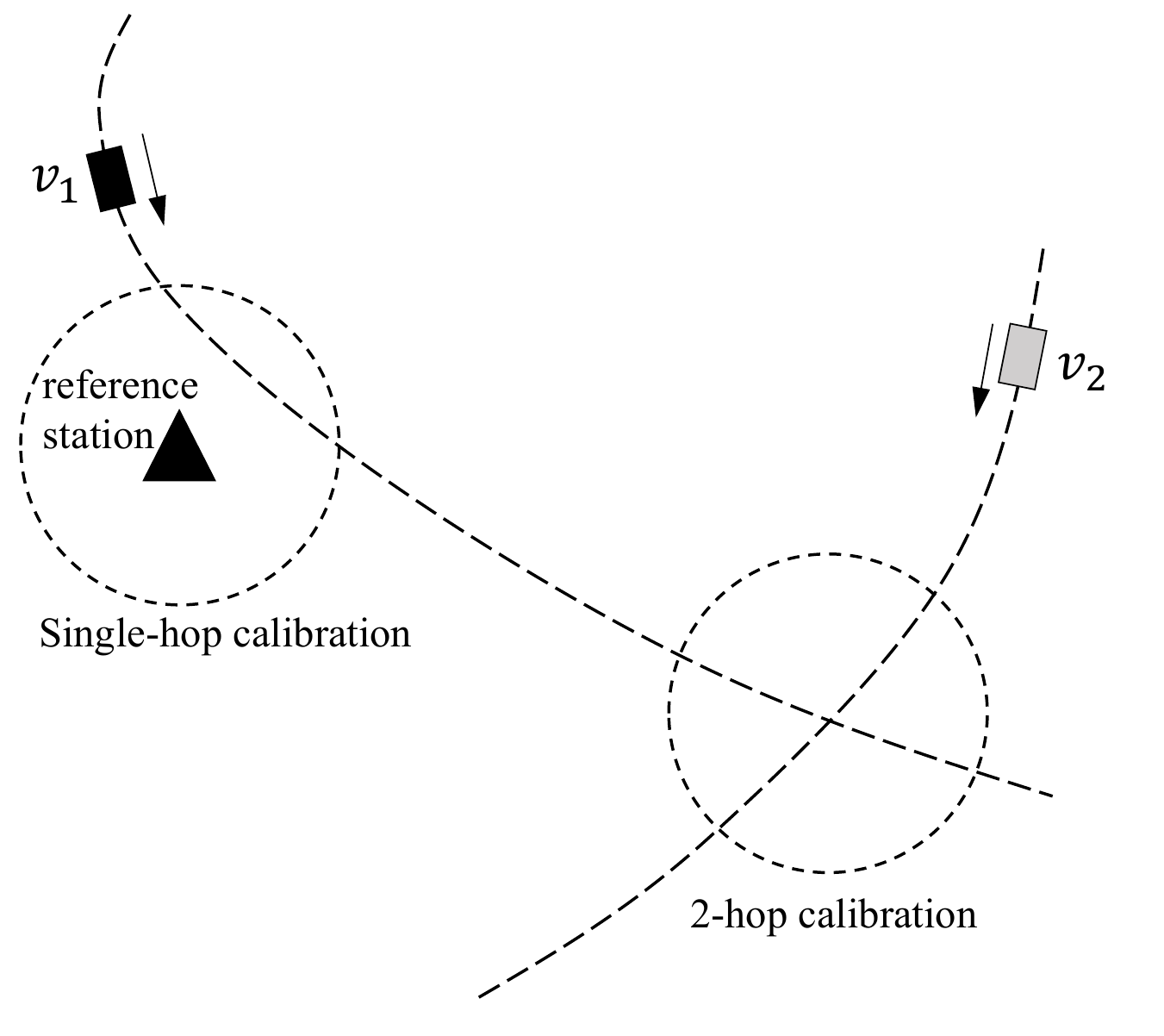}
\caption{Illustration of single- and 2-hop calibration.}
\label{fighop}
\end{figure}

Mobile data calibration via collocation can be categorized as single and multi hop \citep{Hasenfratz2012,FRD2017} as shown in Figure \ref{fighop}. Single-hop calibration forms a data checkpoint when the vehicle $v_1$ is near the reference station $s$, while a 2-hop calibration took place if a calibration path exists starting from the reference station $s$ and the vehicle $v_2$. The cases of k-hop calibration can be similarly defined.

 \cite{Agarwal2020} and \cite{Saukh2012} consider the need for data calibration as a constraint in mobile sensor deployment problems. Another way to address this issue is through the objective function, by either penalizing under-calibrated instances at the expense of non-analytic formulation, or  increasing the spatial weight $w_g$ in \eqref{eqnss} such that more vehicle trips are concentrated where the reference station is located, but without a guarantee that all sensors are uniformly calibrated.

\subsection{Sensor deployment with heterogenous mobility patterns}

The distinct mobility patterns of different types of host vehicles are responsible for their unique capabilities, as well as shortcomings, in certain aspect of drive-by sensing, as illustrated in Figure \ref{figradar}. An issue that has not been systematically explored is the sensing efficacy of a mixed vehicle fleet (e.g. that consisting of taxis, buses and dedicated vehicles). Depending on the level of operational intervention on the host fleet, this issue can be approached as
\begin{itemize}
\item[(1)] A sensor deployment problem, which is an extension of the conceptual model \eqref{conm1}-\eqref{conm5} where the mobility equation \eqref{conm3} is diversified according to the type of vehicles involved. 

\item[(2)] A (partial) operational intervention problem, which aims to maneuver certain types of vehicles to complement the opportunistic sensing offered by other fleets. 
\end{itemize}
While a few studies \citep{FZGJGW2021, Agarwal2020} have considered more than one type of vehicles in the sensing fleet, limited insights are generated regarding the applicability and limitations of such combinations in general, which is worth of further and thorough investigation.

\section{Conclusion and outlook}\label{secconclu}

Urban drive-by sensing (DS) is an emerging data collection paradigm that leverages vehicle mobilities to perform pervasive sensing at low costs. DS has been seen in numerous smart city applications such as air quality sensing, traffic state monitoring, waster disposal surveillance, and disaster response. For a given urban sensing application with certain budget, the ideal distribution of the sensors in a space-time domain is usually difficult to achieve in reality, due to the physical constraints imposed by the mobility patterns of the host vehicles. Therefore, the essential purpose of optimization, whether regarding fleet selection, sensor allocation, or vehicle maneuver, is to remove those constraints such that the resulting sensing efficacy is engineered towards optimality.

Building on a thorough review of existing literature, this paper has addressed the following questions in detail: 
\begin{itemize}
\item[Q1.] quantifying sensing quality;
\item[Q2.] assessing the sensing capabilities of various fleet types;
\item[Q3.] sensor deployment (allocation) optimization;
\item[Q4.] operations of sensing vehicles.
\end{itemize}
While literature in these regards keep expanding, a few practical issues should be noted:
\begin{enumerate}
\item There does not seem to be an agreement on what is the best method to assess sensing quality, even for a given DS application such as air quality sensing. In fact, such an agreement is unlikely to exist since the goodness of a given sensing coverage is dependent on numerous factors, including but not limited to: sensor specification, data quality control standards, subsequent use of the sensing data, and user-defined sensing priority. 
\item There are certain constraints when it comes to mounting sensors on vehicles. Relevant considerations include: power supply (which renders some electric vehicles unsuitable for safety reasons), physical configuration (large sensors cannot be installed without affecting vehicle operation), and vehicle modification (such as electric wires and gas tubes). These and similar constraints need to be contemplated when evaluating the suitability of vehicle fleets for carrying out certain DS tasks. 
\item For third-party vehicles such as taxis and buses, incentivizing schemes or scheduling attempts need to carefully address the potential impact on their operations. While studies have proposed monetary incentives in various scenarios \citep{ZYSLTXM2016}, it is unclear how they pan out on a human behavior level. User privacy is another important aspect of passive sensing that needs to be addressed in future research. 
\end{enumerate}

In summary, urban drive-by sensing is a promising means of collecting crowdsourced sensing data at relatively low costs. In the development and adoption of DS systems, optimization will keep playing an important role to fully harness their potentials. The inherent transport component, and its relevance to logistics and mechanism design \citep{RCS2020, LDLY2022}, suggest that the development of DS can further benefit from the participation of transportation and operations research communities.

\section*{Acknowledgement}
This work is supported by the National Natural Science Foundation of China through grants 72071163 and 72271206, and the Natural Science Foundation of Sichuan Province through grant 2022NSFSC0474.

\end{document}